\documentclass[aip,preprint]{revtex4-1}
\usepackage{graphicx}

\draft 

\begin{document}


\title{Path integral molecular dynamics simulations for Green's function in a system of identical bosons}

\author{Yunuo Xiong}
\email{xiongyunuo20030409@icloud.com}
\affiliation{College of Science, Zhejiang University of Technology, Hangzhou 31023, China}

\author{Hongwei Xiong}
\email{xionghw@zjut.edu.cn}
\affiliation{College of Science, Zhejiang University of Technology, Hangzhou 31023, China}

\begin{abstract}
Path integral molecular dynamics (PIMD) has been successfully applied to perform simulations of large bosonic systems in a recent work (Hirshberg et al., PNAS, 116, 21445 (2019)).
In this work we extend PIMD techniques to study Green's function for bosonic systems. We demonstrate that the development of the original PIMD method enables us to calculate Green's function and extract momentum distribution from our simulations. We also apply our method to systems of identical interacting bosons to  study Berezinskii-Kosterlitz-Thouless transition around its critical temperature.
\end{abstract}


\maketitle

\section{Introduction}
When we study a system of indistinguishable particles, quantum symmetry under permutations of identical particles plays a key role in its behaviors. Therefore, at sufficiently low temperatures, an accurate simulation method should take such symmetries into account. In path integral formulation of quantum mechanics \cite{feynman, kleinert}, exchange effects between identical particles can be automatically taken into account \cite{Tuckerman}. After taking into account the exchange effects, we may construct an equivalent classical system in which every particle corresponds to a different ring polymer \cite{chandler,Parrinello,Miura,Hirshberg,Cao,Cao2,Jang2,Ram,Poly, Craig,Braa,Haber,Thomas} composed of $P$ beads connected by harmonic springs for each permutation of the particles, the contributions from all permutations are weighted equally in the classical partition function.
This equivalent classical system for the partition function of the quantum system may be sampled directly through Monte Carlo method \cite{CeperRMP,boninsegni1,boninsegni2,Dornheim}. 
This classical system may also be simulated using traditional molecular dynamics methods \cite{Tuckerman,Hirshberg} and various physical quantities of interest (such as energies and densities) may be extracted from appropriate estimators. The exact results are recovered in the limit as $P$ goes to infinity (of course, in practice, a finite number of beads is usually good enough to achieve convergence).
\par
Specifically, by dividing the imaginary time into $P$ slices, in a recent pioneering work, Hirshberg, Rizzi and Parrinello \cite{Hirshberg} found that the partition function corresponding to a system of $N$ identical bosons is given by 
\begin{equation}
Z_B^{(N)}=\left(\frac{mP}{2\pi\hbar^2\beta}\right)^{PdN/2} \int e^{-\beta(V_B^{(N)}+\frac{1}{P}U)}d\mathbf R_1...d\mathbf R_N,
\label{partition}
\end{equation}
where $\mathbf R_i$ represents the collection of ring polymer coordinates $(\mathbf r_i^1,...,\mathbf r_i^P)$ corresponding to the $i$th particle. $\beta=1/k_BT$ with $k_B$ the Boltzmann constant and $T$ the system temperature. The system under consideration has $d$ spatial dimensions. $V_B^{(N)}$ considers exchange effects by describing all the possible ring polymer configurations. $U$ is the interaction between different particles, which is given by
\begin{equation}
U = \sum_{l=1}^P V(\mathbf r_1^l,...,\mathbf r_N^l).
\end{equation}
Here $V$ denotes the interaction potential. The expression of $V_B^{(N)}$ is the key to consider the exchange effects of identical bosons, which will be introduced in due course.

The Green's function has become generally recognized as a powerful tool for studying complex quantum many-particle interacting systems \cite{Mahan,Fetter}. It is clear that the extension of  the path integral molecular dynamics (PIMD) \cite{Hirshberg} to the calculation of the Green function would be a key advance in the development of the PIMD method for identical bosons. For example, we may calculate the momentum distribution based on Green's function, while the momentum distribution is an important measurement result in numerous quantum systems such as cold atom experiments \cite{Dalfovo,Anderson,Davis}.

In this paper we develop the PIMD method for calculating the Green's function of the bosonic system. To test our theory and algorithm, we calculate the Green's function of a two-dimensional interacting bosonic system and from it we can infer the existence of the Berezinskii-Kosterlitz-Thouless (BKT) transition \cite{Kosterlitz,Hadzibabic}. Based on the off-diagonal long-range order of the Green's function, we obtain the slow power law  $r^{-1/4}$ at the critical temperature for the BKT transition. We 
also calculate the momentum distribution based on the Green's function. Although it is not the purpose of the present work to reveal new physics, this work paves the way to numerically study the phase transition of many quantum bosonic systems with path integral molecular dynamics and Green's function.

\section{PIMD for Bosons}
Firstly, we give a brief introduction of the method \cite{Hirshberg} by Hirshberg et al., to get the expression of the partition function for identical bosons with classical ring polymer systems.

In the case of distinguishable particles, $V_B^{(N)}$ is just the sum of the individual spring energy of fully connected ring polymer corresponding to each particle. That is,
\begin{equation}
V_B^{(N)}=\frac{1}{2}m\omega_P^2\sum_{l=1}^N\sum_{j=1}^P(\mathbf r_l^{j+1}-\mathbf r_l^j)^2
\label{VBN}
\end{equation}
for distinguishable particles. Here $\omega_P=\sqrt{P}/\beta\hbar$ and $\mathbf r_l^{P+1}=\mathbf r_l^1$. 

For indistinguishable bosons, however, all the possible ring polymer configurations must be summed over in $V_B^{(N)}$ to account for exchange symmetry. In general, there are $N!$ permutations of particles, each corresponds to one ring polymer configuration. 
Even if one considers only the nonequivalent ring polymer configurations there are still exponentially many of them (for $N=3$ there are $3$ nonequivalent configurations). This unfavorable scaling prevents straightforward application of Eq. (\ref{partition}) to simulate the system, since evaluating $V_B^{(N)}$ directly would take exponentially many steps in the number of particles $N$. 

Fortunately, in a recent paper \cite{Hirshberg}, it was shown that $V_B^{(N)}$ may be calculated recursively using the following formula:
\begin{equation}
e^{-\beta V_B^{(\alpha)}}=\frac{1}{\alpha}\sum_{k=1}^\alpha e^{-\beta(E_\alpha^{(k)}+V_B^{(\alpha-k)})},
\label{recursion}
\end{equation}
where $V_B^{(0)}=0$ and $E_\alpha^{(k)}$ is given by
\begin{equation}
E_\alpha^{(k)}(\mathbf R_{\alpha-k+1},...,\mathbf R_\alpha)=\frac{1}{2}m\omega_P^2\sum_{l=\alpha-k+1}^\alpha\sum_{j=1}^P(\mathbf r_l^{j+1}-\mathbf r_l^j)^2,
\end{equation}
where $\mathbf r_l^{P+1}=\mathbf r_{\alpha-k+1}^1$ if $l=\alpha$ and $\mathbf r_l^{P+1}=\mathbf r_{l+1}^1$ otherwise. $E_\alpha^{(k)}$ may be interpreted as partial configuration of particles in which the last $k$ particles are fully connected. In particular, $E_N^{(N)}$ represents the longest ring configuration for $N$ particles. 
In figure 2 of the paper by Hirshberg et al.\cite{Hirshberg},  a beautiful illustration of the idea of the recursion relation has been given.

From the above recursion relation, we may get
\begin{equation}
V_B^{(1)}\rightarrow V_B^{(2)}\rightarrow \cdots  V_B^{(\alpha)}\cdots \rightarrow V_B^{(N)}.
\end{equation}

To see that the above equation for $V_B^{(N)}$ works, we observe that most ring polymer configurations may be split into several interconnected rings which cannot be split further. Assuming $\exp(-\beta V_B^{(N-k)})$ correctly counts all the partial contributions from $N-k$ particles, $\exp(-\beta (E_N^{(k)}+V_B^{(N-k)}))$ then represents all the full configurations constructed by appending another ring of length $k P$ to those from $\exp(-\beta V_B^{(N-k)})$. The only configuration which cannot be represented in this way is that of the longest configuration comprised of $N$ connected particles, this configuration is taken care of in the last term of the above sum $\exp(-\beta E_N^{(N)})$.
\par
Therefore we see that the recursive formula allows us to compute contributions from all ring polymer configurations correctly. Moreover, it is easy to see that evaluating $V_B^{(N)}$ only takes $O(PN^3)$ time, an exponential speedup over simply summing all configurations. To perform molecular dynamics we need the gradient of $V_B^{(N)}$, which may also be computed recursively from Eq. (\ref{recursion}) as
\begin{equation}
-\nabla_{\mathbf r_l^j}V_B^{(\alpha)}=-\frac{\sum_{k=1}^\alpha\left[\nabla_{\mathbf r_l^j}E_\alpha^{(k)}+\nabla_{\mathbf r_l^j}V_B^{(\alpha-k)}\right]e^{-\beta(E_\alpha^{(k)}+V_B^{(\alpha-k)})}}{\sum_{k=1}^\alpha e^{-\beta(E_\alpha^{(k)}+V_B^{(\alpha-k)})}},
\end{equation}
where $\nabla_{\mathbf r_l^j}E_\alpha^{(k)}$ is
\begin{equation}
\nabla_{\mathbf r_l^j}E_\alpha^{(k)}=m\omega_P^2\left[(\mathbf r_l^j-\mathbf r_l^{j+1})+(\mathbf r_l^j-\mathbf r_l^{j-1})\right],
\end{equation}
with the boundary conditions $\mathbf r_l^{P+1}=\mathbf r_{\alpha-k+1}^1$ if $l=\alpha$ and $\mathbf r_l^{P+1}=\mathbf r_{l+1}^1$ otherwise; $\mathbf r_l^{0}=\mathbf r_{\alpha}^P$ if $l=\alpha-k+1$ and $\mathbf r_l^{0}=\mathbf r_{l-1}^P$ otherwise. If $l$ is outside the interval $[\alpha-k+1,\alpha]$ then $\nabla_{\mathbf r_l^j}E_\alpha^{(k)}=0$. By recursive calculation from $\alpha=1$ to $\alpha=N$, we may numerically calculate $-\nabla_{\mathbf r_l^j}V_B^{(N)}$ for molecular dynamics.
\par
The above equations completely define a molecular dynamics algorithm for bosons which scales as $O(PN^3)$, enabling the application of PIMD to large bosonic system. In order to extract physical quantities from our simulations we use various estimators. For example, from the following equation
\begin{equation}
\left<E\right>=-\frac{1}{Z_B^{(N)}}\frac{\partial Z_B^{(N)}}{\partial\beta},
\end{equation}
the energy estimator is given by
\begin{equation}
\left<E\right>=\frac{PdN}{2\beta}+\frac{\left<U\right>}{P}+\left<V_B^{(N)}+\beta\frac{\partial V_B^{(N)}}{\partial\beta}\right>.
\end{equation}
$V_B^{(N)}+\beta\frac{\partial V_B^{(N)}}{\partial\beta}$ may be evaluated as
\begin{equation}
V_B^{(N)}+\beta\frac{\partial V_B^{(N)}}{\partial\beta}=\frac{\sum_{k=1}^N\left[V_B^{(N-k)}+\beta\frac{\partial V_B^{(N-k)}}{\partial\beta}-E_N^{(k)}\right]e^{-\beta(E_N^{(k)}+V_B^{(N-k)})}}{\sum_{k=1}^Ne^{-\beta(E_N^{(k)}+V_B^{(N-k)})}},
\end{equation}
with $V_B^{(0)}+\beta\frac{\partial V_B^{(0)}}{\partial\beta}=0$.
\par
The density estimator is simply given by
\begin{equation}
\rho(\textbf{x})=\left<\frac{1}{P}\sum_{j=1}^P\sum_{k=1}^N\delta(\mathbf r_k^j-\textbf{x})\right>.
\end{equation}
The estimator for the density-density correlation function is
\begin{equation}
\rho(\textbf{x},\textbf{y})=\left<\frac{1}{P^2}\left(\sum_{j=1}^P\sum_{k=1}^N\delta(\mathbf r_k^j-\textbf{x})\sum_{l=1}^P\sum_{n=1}^N\delta(\mathbf r_n^l-\textbf{y})\right)\right>.
\label{DensityC}
\end{equation}

\section{PIMD for thermal Green's Function}
In its current form, PIMD \cite{Hirshberg} is unable to infer thermal Green's function from simulation. The thermal Green's function is defined as
\begin{equation}
G(\textbf{x},\tau_1;\textbf{y},\tau_2)=\left<\mathcal T \left\{\hat \psi(\mathbf y,\tau_2)  \hat \psi^\dagger(\mathbf x,\tau_1)\right\}\right>,
\end{equation}
where $\left<\cdots\right>$ denotes thermal average, $\mathcal T$ is the time-ordering operator.  In addition,
\begin{equation}
\hat \psi(\mathbf x,\tau)=e^{ \hat H \tau}\hat\psi (\mathbf x)e^{-\hat H \tau}, \hat \psi^{\dagger}(\mathbf x,\tau)=e^{ \hat H \tau}\hat\psi^\dagger  (\mathbf x)e^{-\hat H \tau}.
\end{equation}

Let us consider the case $\tau_1<\tau_2$. In this case, we have
\begin{equation}
G(\textbf{x},\tau_1;\textbf{y},\tau_2)=\frac{Tr\left(e^{-\beta_3\hat H}\hat \psi(\mathbf y)  e^{-\beta_2\hat H}\hat \psi^\dagger(\mathbf x)e^{-\beta_1\hat H}\right)}{Tr(e^{-\beta\hat H})},
\end{equation}
with
\begin{equation}
\beta_1=\tau_1,\beta_2=\tau_2-\tau_1,\beta_3=\beta-\tau_2.
\end{equation}

In order to see how we would modify the formulation for PIMD to accommodate Green's function, let us begin by defining the partition function for Green's function $Z_G$ as \cite{boninsegni1,boninsegni2}
\begin{equation}
Z_G=\int d\mathbf xd\mathbf y Tr\left(e^{-\beta\hat H}\mathcal T\left\{\hat\psi(\mathbf y,\tau_2) \hat\psi^\dagger(\mathbf x,\tau_1)\right\}\right).
\end{equation}
When this partition function for Green's function is applied, it means that it should be applied for the case of $G(\textbf{x},\tau_1;\textbf{y},\tau_2)\geq 0$, the same as the situation of diagrammatic Monte Carlo method to calculate the Green's function. Fortunately, in most cases we consider, the condition of $G(\textbf{x},\tau_1;\textbf{y},\tau_2)\geq 0$ is satisfied.

We also define the following normalized bosonic position eigenstates that are symmetric under all permutations of particle labels,
\begin{equation}
\left|N_B\right>\equiv \left|\left\{\mathbf r_1,\cdots,\mathbf r_N\right\}\right>_B=\frac{1}{\sqrt{N!}}\sum_{p\in S_N}\left|p\left\{\mathbf r_1,\cdots,\mathbf r_N\right\}\right>.
\end{equation}
Here $S_N$ represents the set of $N!$ permutation operations.

Of course, we have the following identity operator
\begin{equation}
\hat I_B=\frac{1}{N!}\int d\mathbf r_1\cdots d\mathbf r_N \left|N_B\right> \left<N_B\right |.
\end{equation}
Without considering the exchange symmetry, we also have another form of the identity operator
\begin{equation}
\hat I=\int d\mathbf r_1\cdots d\mathbf r_N \left|\left\{\mathbf r_1,\cdots,\mathbf r_N\right\}\right> \left<\left\{\mathbf r_1,\cdots,\mathbf r_N\right\} \right |.
\end{equation}

We have
\begin{equation}
Z_G=\int d\mathbf xd\mathbf y Tr\left(e^{-\beta\hat H}\mathcal T\left\{\psi(\mathbf y,\tau_2) \psi^\dagger(\mathbf x,\tau_1)\right\}\hat I_B\right).
\end{equation}

For the case $\tau_1<\tau_2$, from the above expression, $Z_G$ can be expressed as
\begin{equation}
Z_G=\frac{1}{N!}\sum_p\int d\mathbf xd\mathbf y \int d\mathbf r_1\cdots d\mathbf r_N \left<p\{\mathbf r\}\right| e^{-\beta_3\hat H}\hat \psi(\mathbf y)  e^{-\beta_2\hat H}\hat \psi^\dagger(\mathbf x)e^{-\beta_1\hat H}\left|\{\mathbf r\}\right>.
\end{equation}

On the other hand, it is easy to prove that 
\begin{equation}
e^{-\beta \hat H}\hat I_B=e^{-\frac{\beta }{P}\hat H}\hat I e^{-\frac{\beta }{P}\hat H} \hat I\cdots \hat I e^{-\frac{\beta }{P}\hat H}\hat I_B.
\end{equation}
It is this equation that facilitates simplification of the derivation of the expression for the partition function in previous section. Here, we also use this equation to derive the recursion relation for $Z_G$.

We divide $\beta$ into $P$ intervals (let us denote the interval length by $\Delta\beta=\beta/P$) and expand each of the evolution operators into products of the form
\begin{equation}
e^{-\beta_1\hat H}=e^{-\Delta\beta \hat H}\cdots e^{-\Delta\beta \hat H}.
\end{equation}

\begin{figure}[htbp]
\begin{center}
 \includegraphics[width=0.75\textwidth]{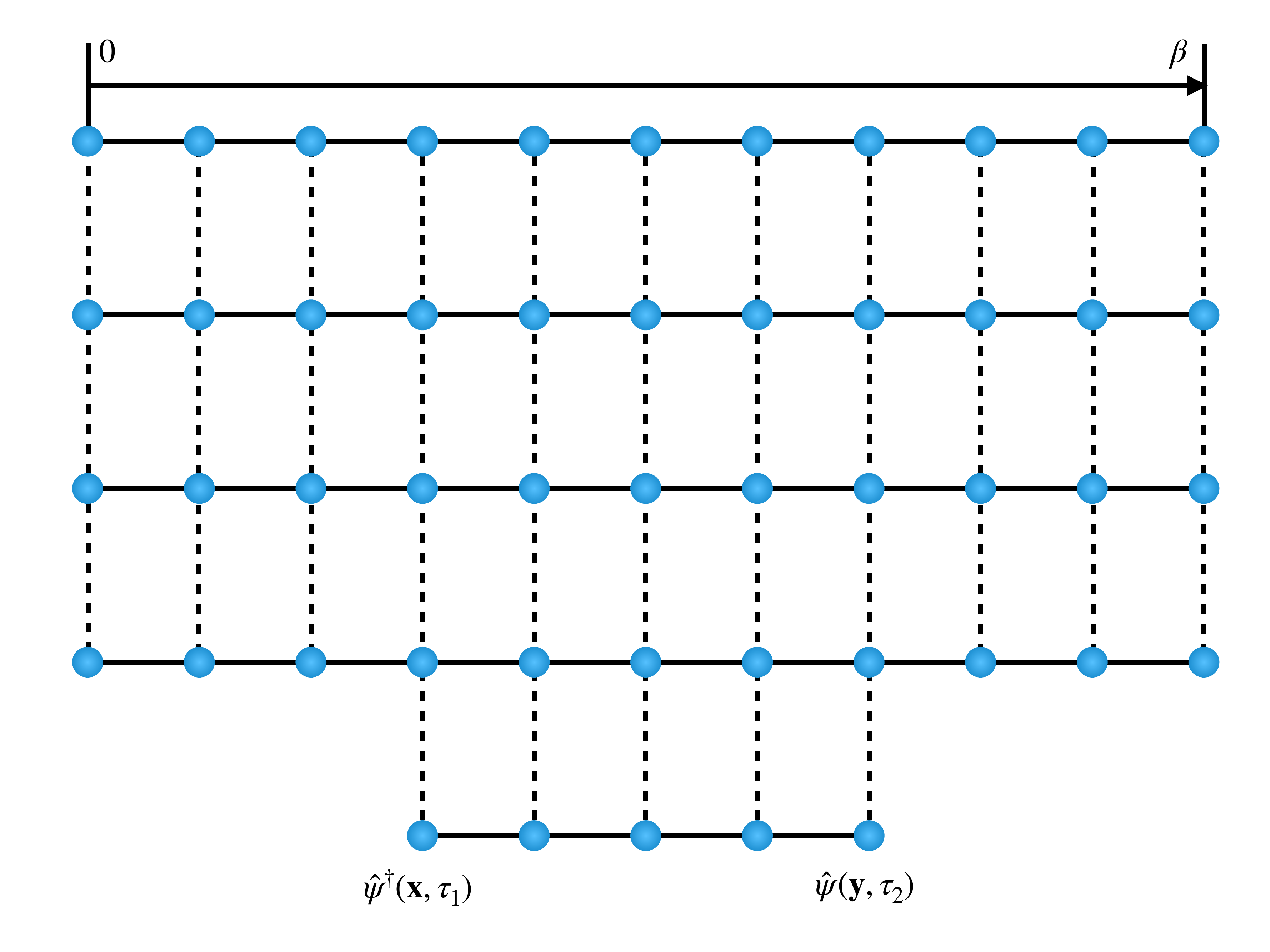} 
\caption{\label{Green1} Illustration of the modified ring polymers for Green's function, for the case $\tau_1<\tau_2$. A new row of beads was added after the $N$th particle, from $\tau_1$ to $\tau_2$.}
\end{center}
\end{figure}

\par
\begin{figure}[htbp]
\begin{center}
 \includegraphics[width=0.75\textwidth]{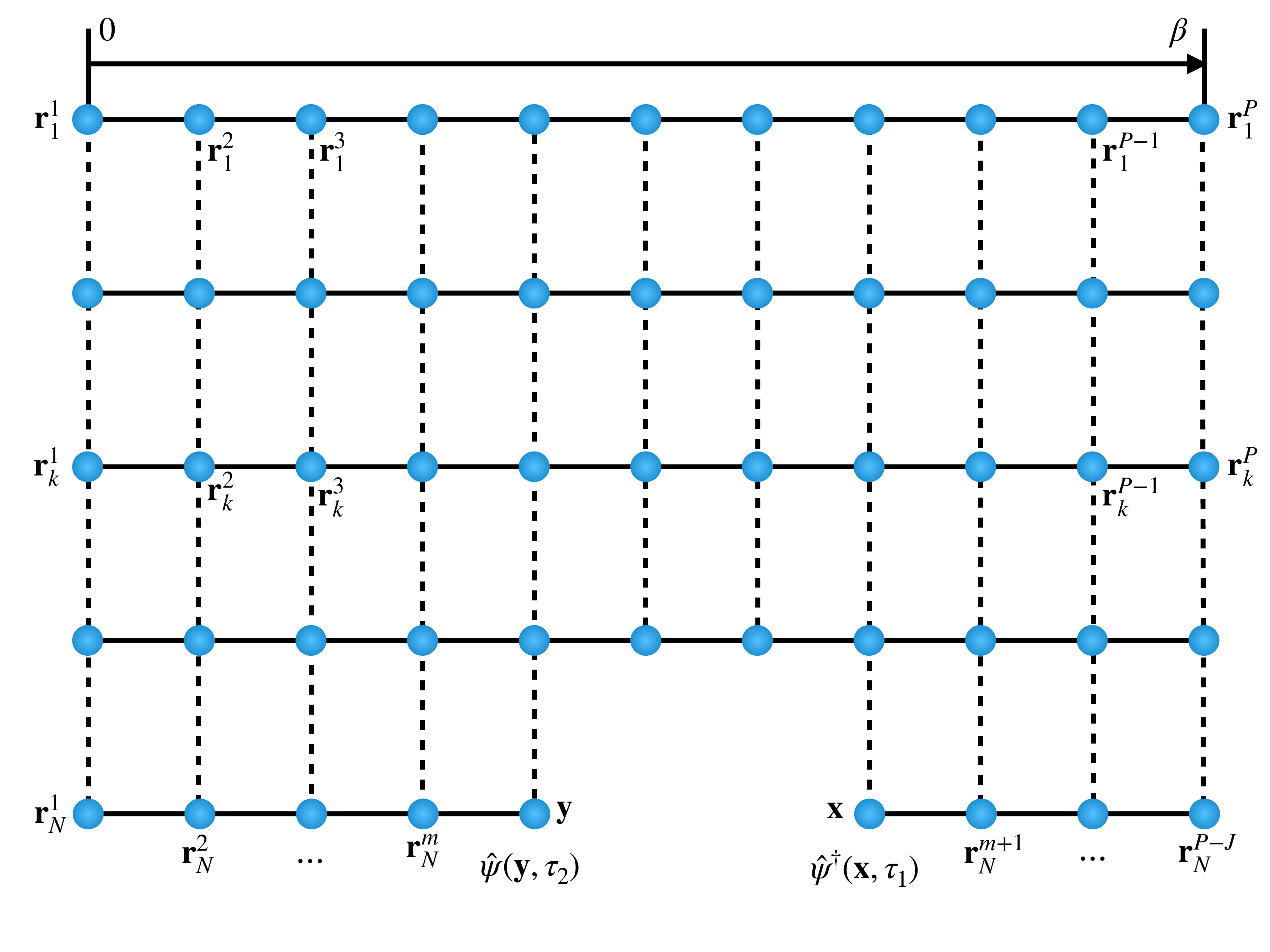} 
\caption{\label{Green2} Modified ring polymers for Green's function, for the case $\tau_1>\tau_2$. A segment of beads was removed from the last particle, from $\tau_2$ to $\tau_1$. The indices of coordinates corresponding to the beads are used in the definition of $E_N^{(k)}$. In particular, the coordinates of the beads at the ends of the gap are $\mathbf y$ and $\mathbf x$, there are no harmonic springs between those two beads.}
\end{center}
\end{figure}

From time $0$ to $\tau_1$, there are ring polymer configurations corresponding to $N$ particles and $P\tau_1/\beta$ beads, just like conventional PIMD. However, from $\tau_1$ to $\tau_2$, because of the presence of $\psi^\dagger(\mathbf x,\tau_1)$, a new particle was created, resulting in $N+1$ particles and $P(\tau_2-\tau_1)/\beta$ beads for this segment. Finally, at $\tau_2$ this new particle will be annihilated by $\psi(\mathbf y,\tau_2)$, leaving us with $N$ particles and $P(\beta-\tau_2)/\beta$ beads for this final segment. So in total, there are $PN+P(\tau_2-\tau_1)/\beta$ particles in molecular dynamics simulations. This is illustrated graphically in Fig. \ref{Green1}. In this figure and Fig. \ref{Green2}, the solid line shows the connection of the beads for the same particle, while the dashed line shows the interparticle interaction between particles at the same imaginary time.
\par

\begin{figure}[htbp]
\begin{center}
 \includegraphics[width=0.75\textwidth]{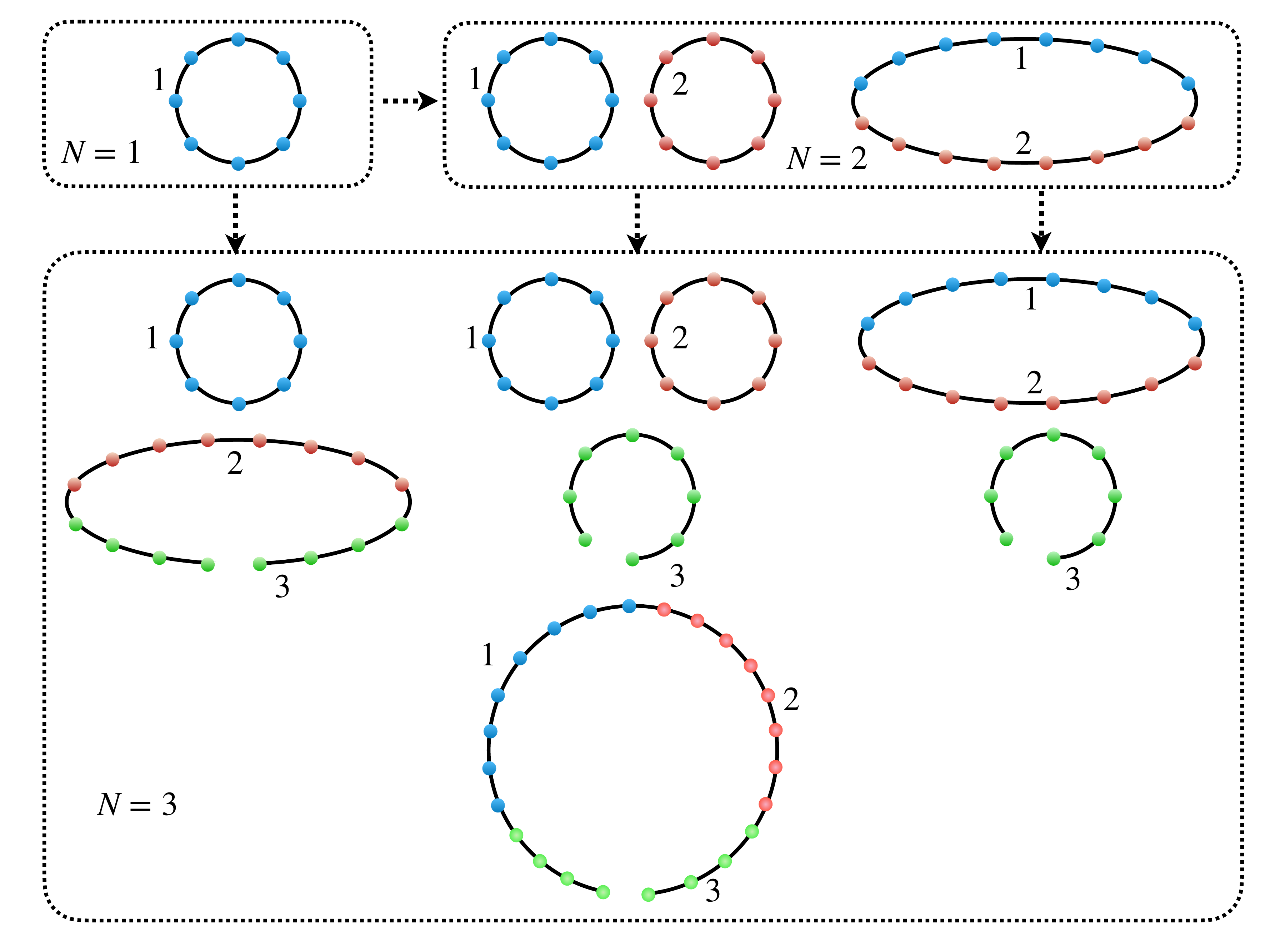} 
\caption{\label{Green3} Illustration of all the modified ring polymer configurations for 3 particles for the case $\tau_1>\tau_2$, to calculate the Green's function. Different colors have been used to distinguish different ring polymers, green stands for the third ring polymer. The gap indicates that a segment of the third ring polymer is missing.}
\end{center}
\end{figure}

The case for $\tau_1>\tau_2$ is similar, except that from $\tau_2$ to $\tau_1$ there will be $N-1$ particles instead of $N+1$. In total there will be $PN-P(\tau_1-\tau_2)/\beta$ beads in our simulations. This situation is shown in Fig. \ref{Green2}, while the role of particle permutations is illustrated in Fig. \ref{Green3} for three identical bosons as an example. In fact, because of periodic boundary along the imaginary time axis (i.e. a cylindrical structure), the case $\tau_1<\tau_2$ may be converted to $\tau_1>\tau_2$ by regarding the new particle as being annihilated from time $\tau_2$ to $\tau_1$, rather than created from $\tau_1$ to $\tau_2$. 

The case of $\tau_1>\tau_2$ is easier to implement so we will work with it.
\par
The partition function of the Green's function now takes the form
\begin{equation}
Z_G=\left(\frac{mP}{2\pi\hbar^2\beta}\right)^{(PN-J)d/2} \int d\mathbf x d\mathbf y\int d\mathbf R_1\cdots d\mathbf R_{N-1}d\mathbf R_N e^{-\beta (V_G^{(N)}+\frac{1}{P}U_G)}.
\label{PartitionZG}
\end{equation}
Here $\mathbf R_i(1\leq i\leq N-1)$ represents the collection of ring polymer coordinates ($\mathbf r_i^1,\cdots,\mathbf r_i^P$) corresponding to the $i$th particle. $\mathbf R_N$ represents $(\mathbf r_N^1,\cdots, \mathbf r_N^l, \mathbf y,\mathbf x,\mathbf r_N^{l+1},\cdots, \mathbf r_N^{P-J})$, with $l=\tau_2/\Delta \beta-1$. $J$ is given by (rounded appropriately)
\begin{equation}
\frac{\tau_1-\tau_2}{\beta}=\frac{J}{P}.
\end{equation}
\par
In order to modify our PIMD program to accommodate this change, it suffices to modify $E_N^{(k)}$ and its gradient. The interaction potential $U_G$ between particles can also be modified straightforwardly. $U_G$ consists of the summation of interaction between beads connected by dashed lines in Fig. 2. 
The recursion relation for $V_G^{(\alpha)}$ is
\begin{equation}
e^{-\beta V_G^{(\alpha)}}=\frac{1}{\alpha}\sum_{k=1}^\alpha e^{-\beta(E_\alpha^{(k)}+V_G^{(\alpha-k)})},
\label{recursion2}
\end{equation}
where $V_G^{(0)}=0$. From the above recursion relation, we may get $V_G^N$ needed in Eq. (\ref{PartitionZG}). One may refer to the code in GitHub to understand how the algorithm implements $V_G^N$ and $U_G$.

Let us denote positions of the beads corresponding to the $N$th particle at times $\tau_1$ and $\tau_2$ as $\mathbf x$ and $\mathbf y$ respectively, we see that for the $N$th particle, $E_N^{(N)}$ should be modified as follows

\begin{eqnarray*}
E_N^{(N)}=&&\frac{1}{2}m\omega_P^2\left\{\sum_{l=1}^{N-1}\sum_{j=1}^P(\mathbf r_l^{j+1}-\mathbf r_l^j)^2\right.\\
&&+(\mathbf r_N^2-\mathbf r_N^1)^2+...+(\mathbf y-\mathbf r_N^m)^2\\
&&\left.+(\mathbf r_N^{m+1}-\mathbf x)^2+...+(\mathbf r_N^{P-J+1}-\mathbf r_N^{P-J})^2\right\},
\end{eqnarray*}
where the boundary conditions remain unaltered (with $\mathbf r_N^{P-J+1}$ now equal to $\mathbf r_1^1$). The formula for the general $E_\alpha^{(k)}$ is the same as before when $\alpha<N$; when $\alpha=N$, it includes all the usual spring energies between beads but without the $(\mathbf x-\mathbf y)^2$ term. 
 
The gradient of $E_N^{(k)}$ should be modified accordingly. For example, the gradient of $E_N^{(N)}$ with respect to $\mathbf x$ is
\begin{equation}
\nabla_{\mathbf x}E_N^{(N)}=m\omega_P^2(\mathbf x-\mathbf r_N^{m+1}).
\end{equation}

With the above changes, Green's function can be estimated as
\begin{equation}
G(\textbf{x}',\tau_1;\textbf{y}',\tau_2)=\left<\delta(\textbf{x}-\textbf{x}')\delta(\textbf{y}-\textbf{y}')\right>.
\end{equation}
where $\textbf{x}$ and $\textbf{y}$ denote the positions of the two beads at the end of the gap in Fig. \ref{Green2}. It is clear that the Green's function is different from the density-density correlation function (\ref{DensityC}).
The density-density correlation function should be calculated based on the distribution given by Eq. (\ref{partition}), while the Green's function is calculated based on the distribution given by  Eq. (\ref{PartitionZG}).

From Green's function, we may get the momentum distribution of the system. Assuming that $\hat a({\textbf p})$ and 
$\hat a^\dagger({\textbf p})$ are the annihilation and creation operators for a particle with momentum $\textbf p$, we have
\begin{equation}
\hat \psi (\textbf x)=\frac{1}{(2\pi\hbar)^{d/2}}\int d\textbf p \hat a({\textbf p})e^{\frac{i}{\hbar}\textbf p\cdot\textbf x},
\hat \psi^\dagger (\textbf x)=\frac{1}{(2\pi\hbar)^{d/2}}\int d\textbf p \hat a^\dagger({\textbf p})e^{-\frac{i}{\hbar}\textbf p\cdot\textbf x}.
\end{equation}
Here $d$ is the spatial dimension of the system. From the above expression, we may also get
\begin{equation}
\hat a({\textbf p})=\frac{1}{(2\pi\hbar)^{d/2}}\int d\textbf x \hat \psi (\textbf x) e^{-\frac{i}{\hbar}\textbf p\cdot\textbf x},
\hat a^\dagger({\textbf p})=\frac{1}{(2\pi\hbar)^{d/2}}\int d\textbf x  \hat \psi^\dagger (\textbf x) e^{\frac{i}{\hbar}\textbf p\cdot\textbf x}.
\end{equation}

The momentum density distribution is
\begin{equation}
\rho(\textbf{p})=\frac{Tr(e^{-\beta\hat H}\hat a^\dagger({\textbf p})\hat a({\textbf p}))}{Tr(e^{-\beta\hat H})}.
\end{equation}
After simple derivations, we get
\begin{equation}
\rho(\textbf{p})=\frac{1}{(2\pi\hbar)^d}\int d{\textbf x}d{\textbf y} G({\textbf x},\tau_1;{\textbf y},\tau_2)
e^{\frac{i}{\hbar}\textbf p\cdot({\textbf x}-{\textbf y})}.
\end{equation}
Here $\tau_1=\tau_2+0^+$.

\section{Results}
In order to test our method, we apply our algorithm to study various systems of identical bosons. In all of the following simulations,  we will use massive Nos\'e-Hoover chain \cite{Nose1,Nose2,Hoover,Martyna,Jang} to establish constant temperature for the system, where each degree of freedom of the system has been coupled to a separate Nos\'e-Hoover thermostat. The number of beads used decreases proportionally as temperature increases to ensure numerical stability and assure convergence, so that $\Delta\beta$ is the same for different temperatures. In all of the following we checked convergence with respect to the number of beads and MD steps performed. For details of how to assure the convergence, one may refer to the supplementary material in \cite{Hirshberg}.

For identical bosons in a harmonic trap with frequency $\omega$, we consider Gaussian interaction \cite{Hirshberg,Mujal} between bosons of the form
\begin{equation}
V(\mathbf r_1,...,\mathbf r_N)=\frac{1}{2}\sum_{i\neq j=1}^N\frac{g}{\pi s^2}e^{-\frac{(\mathbf r_i-\mathbf r_j)^2}{s^2}}.
\label{Vinteraction}
\end{equation}
Here $V$ is in unit of $\hbar\omega$, while the length is in unit of $\sqrt{\hbar/m\omega}$. The dimensionless parameter $\tilde \beta=\hbar\omega\beta$.

\begin{figure}[htbp]
\begin{center}
 \includegraphics[width=0.75\textwidth]{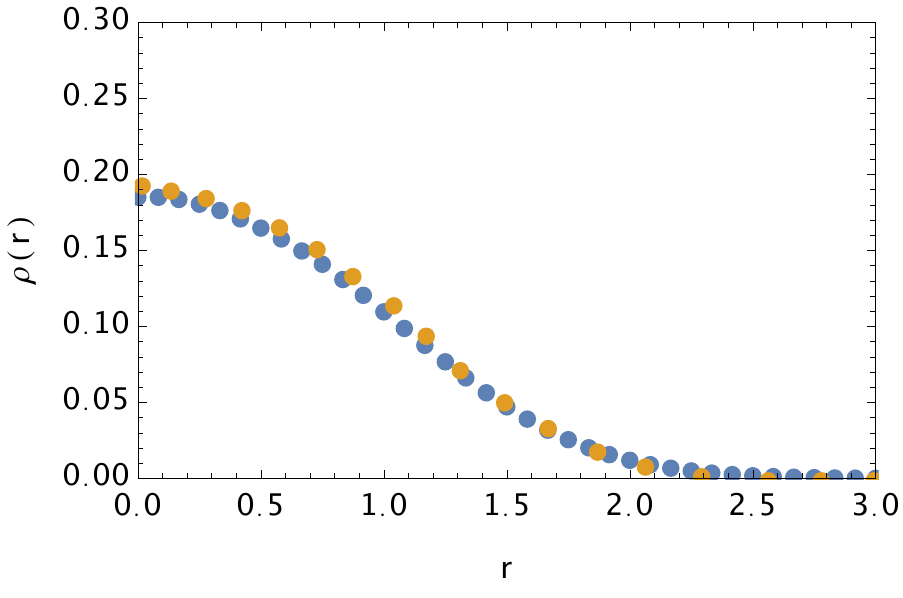} 
\caption{\label{density} The blue is the density function for $4$ particles  ($64$ beads and $10^7$ MD steps), for the case $s=0.5$ and $g=3.0$ at a temperature of $\tilde \beta=6.0$. We do not give the error bar because the statistical fluctuations are negligible. The orange is the data from Fig. 3 of previous work \cite{Hirshberg}, and reasonable agreement is clearly shown.} 
\end{center}
\end{figure}

Firstly, we calculate the density function $\rho$ with the same parameter in the work by Hirshberg, et al. \cite{Hirshberg}, as shown in Fig. \ref{density} to test our calculations. We have checked that our result of the energy and density distribution agree with previous results \cite{Hirshberg}.

\begin{figure}[htbp]
\begin{center}
 \includegraphics[width=0.75\textwidth]{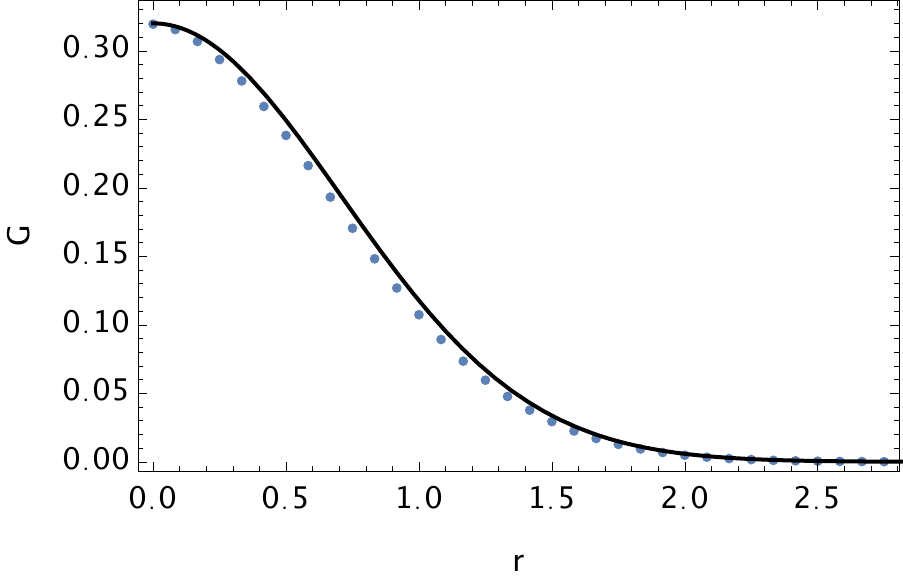} 
\caption{\label{HGreen} Green's function for 8 particles ($64$ beads and $10^7$ MD steps) without interaction in a harmonic trap at low temperature ($\tilde \beta=6.0$). The solid line corresponds to analytical results for ground state bosons. We do not give the error bar because the statistical fluctuations are negligible.}
\end{center}
\end{figure}

Secondly, we calculate the Green's function for  $8$ identical particles in the harmonic trap without interaction. For temperature $k_B T<<\hbar\omega$, the Green's function has the following analytical result:
\begin{equation}
G(\textbf{0},\textbf{r})\simeq N\frac{m\omega}{\pi\hbar}e^{-\frac{m\omega}{2\hbar}(x^2+y^2)}.
\end{equation}
In Fig. \ref{HGreen}, we give the analytical and numerical results, and good agreement is shown. This proves the validity of our method to calculate the Green's function.

Finally, to show the universal BKT transition, we turn to consider the Green's function without the harmonic trap.
With the length unit $a$, the energy unit is $E_u=\hbar^2/2ma^2$. In our numerical calculations, $L$ is the dimensionless length
so that the box is $L\times L$ with periodic boundary condition, while the dimensionless $\tilde\beta$ is defined as $\tilde \beta=\beta E_u$. The interaction potential takes the same form as Eq. (\ref{Vinteraction}), with new length unit $a$ and energy unit $E_u$.

Then, we keep $\tau_1-\tau_2\neq0$ a constant (we used $\tau_1-\tau_2=\beta/3$, or equivalently $J=P/3$) and study the case where $s=0.5$ and $g=3.0$. The existence of interparticle interactions is the necessary condition to observe the BKT transition. We keep the size of simulation box (we used $L=3.0$) and number of particles fixed ($N=8$), and vary the temperature in order to study the BKT transition of the system.
\par

\begin{figure}[htbp]
\begin{center}
 \includegraphics[width=0.75\textwidth]{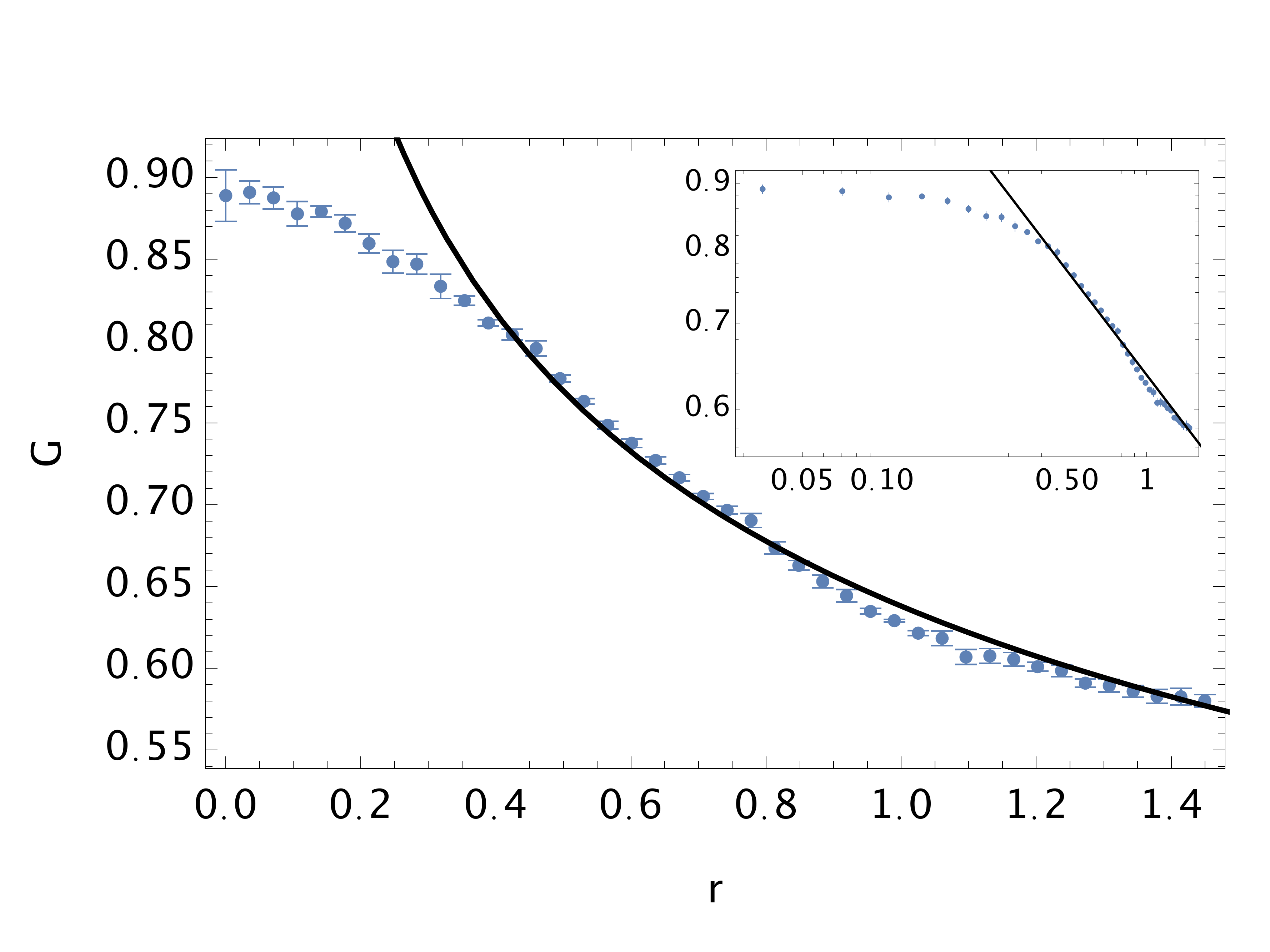} 
\caption{\label{critical}At the critical temperature, Green's function takes the form of a power-law decay function. The critical temperature we found is given by $\tilde\beta^{-1}=1.625$. The inset shows the log-log plot for Green's function at the critical temperature of the BKT transition. Also shown are the fitted line of $0.638r^{-0.27}$. We use the number
of beads $P = 12$ and 5 independent trajectories each consisting of $10^7$ MD steps. The slight difference with $a r^{-1/4}$ law is due to the finite size and finite number of particles in our numerical simulation. Note that the coefficient $a$ is not important for the physics of BKT transition. In addition, in order to fit the analytical result of $r^{-1/4}$ law at small $r$ values, systems with much larger number of particles should be used.}
\end{center}
\end{figure}

\begin{figure}[htbp]
\begin{center}
 \includegraphics[width=0.75\textwidth]{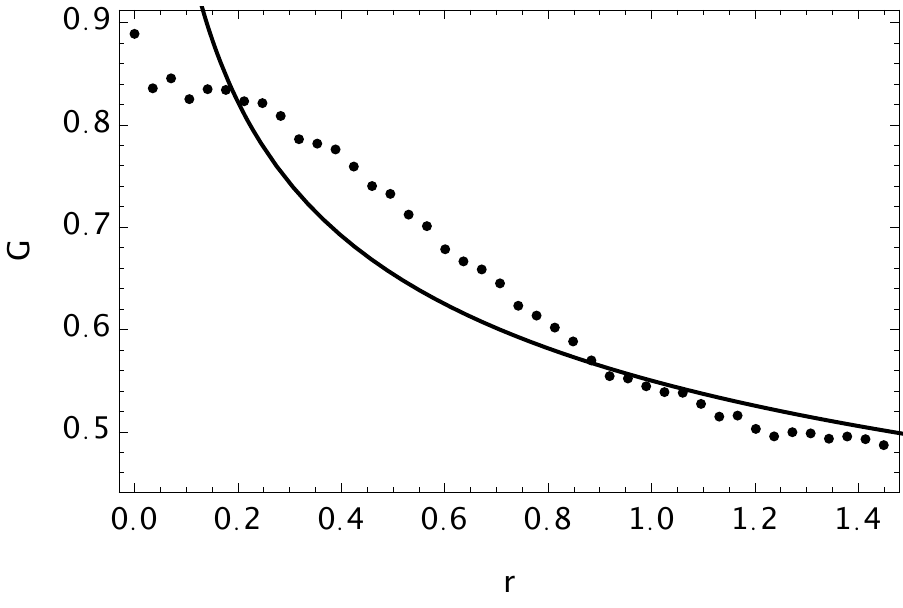} 
\caption{\label{above} At temperatures slightly above the critical temperature, Green's function is Gaussian-like. There are 8 particles (12 beads and $10^7$ MD steps) and the temperature is $\tilde\beta^{-1}=1.75$, the size of the box is $L=3.0$. We also plot the $r^{-1/4}$ law as a comparison.}
\end{center}
\end{figure}

At temperature far below the critical temperature, the Green's function is flat, which deviates significantly from $r^{-1/4}$. However, as we increase the temperature we found that at one specific critical temperature, the Green's function takes the approximate form of the function $r^{-1/4}$, signifying the occurrence of the BKT transition (see Fig. \ref{critical}). In the inset of this figure, we give the log-log plot for the Green's function at the critical temperature of the BKT transition. Since the $r^{-1/4}$ law is for an infinite system in the thermodynamic limit, while we consider here a finite system with periodic boundary condition, we only compare the region $0<r<L/2$ to make the comparison reasonable. Increasing the temperature further, the Green's function quickly becomes Gaussian-like. In Fig. \ref{above}, we give the Green's function for the temperature slightly above the critical temperature.

The $1/4$ power law at the BKT transition temperature is a universal result of the BKT transition, which is not sensitive to the interaction strength or particle number. By calculating the Green's function with different $g$, we do find this universal result. Compared with the Green's function for noninteracting bosons, we notice significant statistical fluctuations which is due to the fluctuations at the phase transition and interparticle interactions.
Compared with the work \cite{Filinov} about the BKT transition in two-dimensional dipole systems without considering the Green's function, it is clear that the calculation of the Green's function provides the chance to reveal more fundamental characteristics of the BKT transition.

\begin{figure}[htbp]
\begin{center}
 \includegraphics[width=0.75\textwidth]{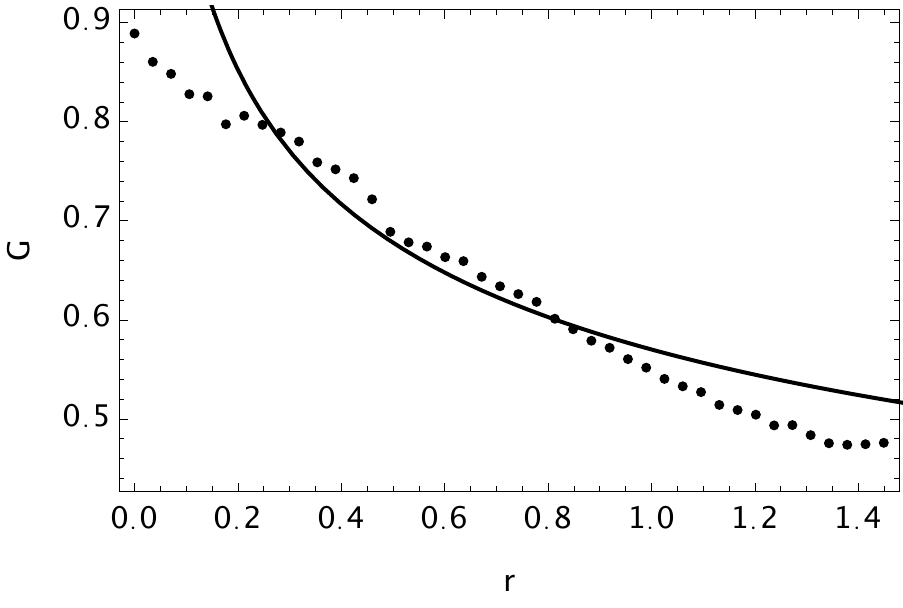} 
\caption{\label{ideal} In a system with no interactions (i.e. $g=0$), no BKT phase transitions were found. The figure illustrates the closest Green's function we could find to the $r^{-1/4}$ law, at a temperature given by $\tilde\beta^{-1}=1.5$ (12 beads and $10^7$ MD steps). The deviation from the $r^{-1/4}$ law is obvious.}
\end{center}
\end{figure}

We also calculate the case where there is no interaction. At low temperature, the Green's function as determined from our simulations is completely flat. As we increase the temperature, the Green's function takes the shape of a flat function plus a Gaussian function. At high temperature, Green's function becomes Gaussian-like. No BKT transition was observed as we vary the temperature in this case.
We found that without interaction, it is impossible to obtain the $r^{-1/4}$  law for the off-diagonal long-range order at any temperature. In Fig. \ref{ideal}, we show the closest Green's function we could find to the function $r^{-1/4}$, which has obvious deviation from the $r^{-1/4}$ law.

\begin{figure}[htbp]
\begin{center}
 \includegraphics[width=0.75\textwidth]{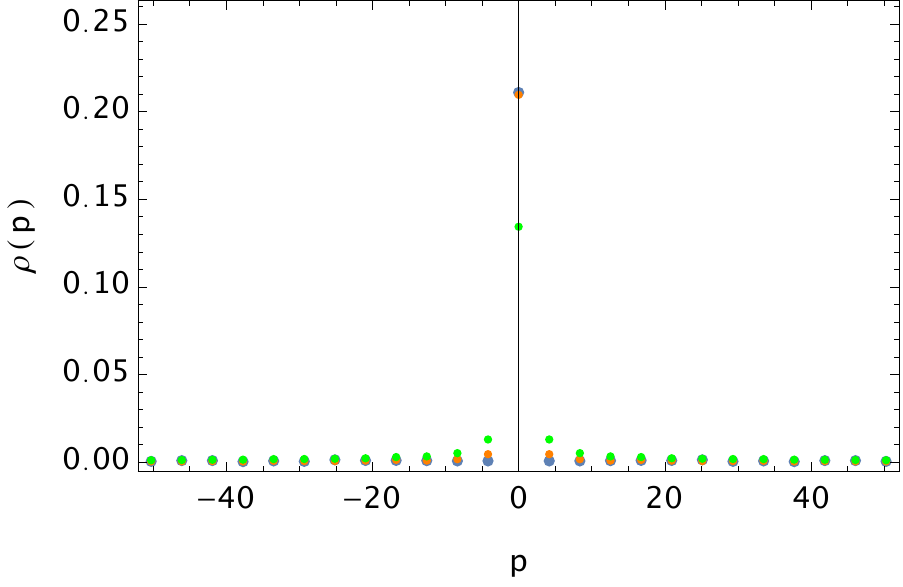} 
\caption{\label{Mdistribution}  The momentum distribution ($10^7$ MD steps) is shown for different temperatures $\tilde \beta^{-1}=1/6, 1, 2$, corresponding to blue, orange and green respectively. We may notice that the width of the momentum distribution decreases with the decreasing of the temperature.}
\end{center}
\end{figure}

From the calculations of the Green's function, we may get the momentum distribution of the system. In Fig. \ref{Mdistribution}, we give the momentum distribution for different temperatures. For a uniform system, we have
\begin{equation}
G(\textbf{x},\tau_1;\textbf{y},\tau_2)=G(\textbf{x}-\textbf{y},\tau_1-\tau_2).
\end{equation}
In this case, the momentum distribution becomes
\begin{equation}
\rho(\textbf p)=\frac{L^d}{(2\pi\hbar)^d}\int d{\textbf x} G({\textbf x},\tau_1-\tau_2)
e^{\frac{i}{\hbar}\textbf p\cdot{\textbf x}},~\tau_1=\tau_2+0^+.
\end{equation}
In Fig. \ref{Mdistribution}, we notice that the width of the momentum distribution decreases with the decreasing of the temperature. However, our simulations  show that it is not easy to calibrate accurately the BKT transition from the momentum distribution in this figure. This is the reason why the momentum distribution of two quasi-2D ultracold atomic gases after an interference of free expansion is measured to demonstrate the BKT transition in cold atom experiments \cite{Hadzibabic}. In this experiment \cite{Hadzibabic}, the $r^{-1/4}$ law in the Green's function is verified after systematic analysis of the observed momentum distribution. In fact, the momentum distribution has a wide range of applications in revealing numerous quantum phenomena \cite{Dalfovo,Anderson,Davis}. 

\section{Conclusions}
As a summary, in the past path integral molecular dynamics has found few applications in the field of identical particles, because of the inefficiencies when evaluating force and potential functions. Recently, it was shown \cite{Hirshberg} how to perform PIMD efficiently, resulting in a polynomial algorithm for the simulations, enabling the application of PIMD to large bosonic systems. The original PIMD methodology can only determine physical quantities such as energy and density from simulations. In this work we showed how to extend PIMD technique to study Green's function for bosonic systems. We applied it to a toy model in order to study its BKT transition and verify the correctness of our method. The technique may be extended to more complex systems to study their phase transition behaviors. In particular, the momentum distribution extracted from the Green's function will have a wide range of applications in cold atom physics, where the momentum distribution is measured in most experiments after a free expansion of the cold atomic gases \cite{Hadzibabic,Dalfovo,Anderson,Davis}. It would be promising to calculate the Green's function for supersolid phase in high-pressure deuterium bosonic system \cite{Deuterium} and general fermionic systems \cite{HirshbergFermi}, which would provide the key information to understand these quantum systems.

\begin{acknowledgments}
This work
is partly supported by the National Natural Science Foundation of China under grant numbers 11175246, and 11334001.
\end{acknowledgments}

\textbf{DATA AVAILABILITY}

The data that support the findings of this study are available from the corresponding author upon reasonable request.
The code of this study is openly available in GitHub (https://github.com/xiongyunuo/PIMD-for-Green-Function).


\begin{thebibliography}{10}

\bibitem{feynman} R. P.~Feynman and A. R.~Hibbs, Quantum mechanics and path integrals, Dover Publications, New York (2010).

\bibitem{kleinert} H.~Kleinert, Path integrals in quantum mechanics, statistics, polymer physics, and financial markets, World Scientific, Singapore (2009).

\bibitem{Tuckerman} M. E.~Tuckerman, Statistical mechanics: theory and molecular simulation, Oxford University, New York (2010).


\bibitem{chandler} D.~Chandler and P. G.~Wolynes, Exploiting the isomorphism between quantum theory and classical statistical mechanics of polyatomic fluids, {\text{J.~Chem.~Phys.}~\textbf{74}, 4078} (1981).

\bibitem{Parrinello} M.~Parrinello and A.~Rahman, Study of an F center in molten KCl, \text{J. Chem. Phys.}~\textbf{80}, 860 (1984).


\bibitem{Miura} S. Miura and S. Okazaki, Path integral molecular dynamics for Bose-Einstein and Fermi-Dirac statistics. \text{J. Chem. Phys.} \textbf{112}, 10116 (2000).

\bibitem{Hirshberg} B. Hirshberg, V. Rizzi, and M. Parrinello, Path integral molecular dynamics for bosons, \text{Proc. Natl. Acad. Sci. U. S. A.}~\textbf{116}, 21445 (2019).


 \bibitem{Cao} J. Cao and G. A. Voth,  The formulation of quantum statistical mechanics based on the Feynman path centroid density. I. Equilibrium properties, \text{J. Chem. Phys.} \textbf{100},  5093 (1994).
 
 \bibitem{Cao2} J. Cao and G. A. Voth, The formulation of quantum statistical mechanics based on the Feynman path centroid density. II. Dynamical properties,  \text{J. Chem. Phys.} \textbf{100}, 5106 (1994).
 
 \bibitem{Jang2} S. Jang and G. A. Voth, A derivation of centroid molecular dynamics and other approximate time evolution methods for path integral centroid variables,  \text{J. Chem. Phys.} \textbf{111}, 2371 (1999).
 
\bibitem{Ram} R. Ram\'iRez and T. L\'oPez-Ciudad, The Schr\"odinger formulation of the Feynman path centroid density, \text{J. Chem. Phys.} \textbf{111}, 3339 (1999).
 
\bibitem{Poly} E. A. Polyakov, A. P.  Lyubartsev, and P. N.  Vorontsov-Velyaminov,  Centroid molecular dynamics: Comparison with exact results for model systems, \text{J. Chem. Phys.} \textbf{133}, 194103 (2010). 

 \bibitem{Craig} I. R. Craig and D. E. Manolopoulos, Quantum statistics and classical mechanics: Real time correlation functions from ring polymer molecular dynamics,  \text{J. Chem. Phys.}  \textbf{121}, 3368 (2004). 
 
 \bibitem{Braa} B. J. Braams and D. E. Manolopoulos, On the short-time limit of ring polymer molecular dynamics, \text{J. Chem. Phys.} \textbf{125}, 124105 (2006).
 
 \bibitem{Haber} S. Habershon, D. E. Manolopoulos, T. E. Markland, and T. F. Miller 3rd, Ring-polymer molecular dynamics: quantum effects in chemical dynamics from classical trajectories in an extended phase space, \text{Annu. Rev. Phys. Chem.} \textbf{64}, 387 (2013).
 
 \bibitem{Thomas} T. E. Markland and M. Ceriotti, Nuclear quantum effects enter the mainstream, \text{Nat. Rev. Chem.} \textbf{2}, 0109 (2018).
 
\bibitem{CeperRMP} D. M. Ceperley, Path integrals in the theory of condensed helium, \text{Rev. Mod. Phys.} \textbf{67}, 279 (1995).

\bibitem{boninsegni1} M.~Boninsegni, N. V.~Prokof’ev, and B. V.~Svistunov, Worm algorithm and diagrammatic Monte Carlo: A new approach to continuous-space path integral Monte Carlo simulations,  {\text{Phys.~Rev.~E}~\textbf{74}, 036701} (2006).

\bibitem{boninsegni2} M.~Boninsegni, N. V.~Prokof’ev, and B. V.~Svistunov, Worm algorithm for continuous-space path integral Monte Carlo simulations,  {\text{Phys.~Rev.~Lett.}~\textbf{96}, 070601} (2006).

\bibitem{Dornheim} T.~Dornheim, The Fermion sign problem in path integral Monte Carlo simulations: quantum dots, ultracold atoms, and warm dense matter, \text{Phys. Rev. E}~\textbf{100}, 023307 (2019).


\bibitem{Mahan} G. D. Mahan, Many-particle physics, Plenum, New York (2000).

\bibitem{Fetter} A. L. Fetter and J. D. Walecka, Quantum theory of many-particle systems, McGraw-Hill, New York (1971).


\bibitem{Dalfovo} F. Dalfovo, S. Giorgini, L. P. Pitaevskii, and S. Stringari, Theory of Bose-Einstein condensation in trapped gases,
Rev. Mod. Phys. \textbf{71}, 463 (1999).

\bibitem{Anderson} M. H. Anderson, J. R. Ensher, M. R. Matthews, C. E. Wieman, and E. A. Cornell, Observation of Bose-Einstein condensation in a dilute atomic vapor, Science \textbf{269}, 198 (1995).

\bibitem{Davis} K. B. Davis et al., Condensation in a gas of sodium atoms, Phys. Rev. Lett. \textbf{75}, 3973 (1995).

\bibitem{Kosterlitz} J. M. Kosterlitz and D. J. Thouless, Metastability and phase transitions in two dimensional systems, \text{J. Phys. C} \textbf{6}, 1181 (1973).

\bibitem{Hadzibabic} Z. Hadzibabic, P. Kr\"uger, M. Cheneau, B. Battelier, and J. Dalibard, Berezinskii-Kosterlitz-Thouless crossover in a trapped atomic gas, \text{Nature} \textbf{41}, 1118 (2006).


\bibitem{Nose1} S. Nos\'e, A molecular dynamics method for simulations in the canonical ensemble, \text{Mol. Phys.} \textbf{52}, 255 (1984).

\bibitem{Nose2} S. Nos\'e, A unified formulation of the constant temperature molecular dynamics methods, \text{J. Chem. Phys.} \textbf{81}, 511 (1984).

\bibitem{Hoover} W. G. Hoover, Canonical dynamics: Equilibrium phase-space distributions, \text{Phys. Rev. A} \textbf{31}, 1695 (1985).

\bibitem{Martyna} G. J. Martyna, M. L. Klein, and M. Tuckerman, Nos\'e-Hoover chains: The canonical ensemble via continuous dynamics, \text{J. Chem. Phys.} \textbf{97}, 2635 (1992).

\bibitem{Jang} S. Jang and G. A. Voth, Simple reversible molecular dynamics algorithms for Nos\'e-Hoover chain dynamics, \text{J. Chem. Phys.}~\textbf{107}, 9514 (1997).

\bibitem{Mujal} P. Mujal, E. Sarl\'e, A. Polls, and B. Juli\'a-D\'az, Quantum correlations and degeneracy of identical bosons in a two-dimensional harmonic trap, \text{Phys. Rev. A} \textbf{96}, 043614 (2017).

\bibitem{Filinov} A. Filinov, N. V. Prokof’ev, and M. Bonitz, Berezinskii-Kosterlitz-Thouless transition in two-dimensional dipole systems, \text{Phys. Rev. Lett.} \textbf{105}, 070401 (2010).

\bibitem{Deuterium}   C. W. Myung, B. Hirshberg, and M. Parrinello, 
Prediction of a supersolid phase in high-pressure deuterium, \text{Phys. Rev. Lett.} \textbf{128}, 045301 (2022).

\bibitem{HirshbergFermi} B. Hirshberg,  M. Invernizzi, and  M. Parrinello, Path integral molecular dynamics for fermions: Alleviating the sign problem with the Bogoliubov inequality, \text{J. Chem. Phys.} \textbf{152}, 171102 (2020).

\end{thebibliography}
\end{document}